\documentclass[sn-mathphys-num]{sn-jnl}% Math and Physical Sciences Numbered Reference Style 
\usepackage{graphicx}%
\usepackage{multirow}%
\usepackage{amsmath,amssymb,amsfonts}%
\usepackage{amsthm}%
\usepackage{mathrsfs}%
\usepackage[title]{appendix}%
\usepackage{xcolor}%
\usepackage{textcomp}%
\usepackage{manyfoot}%
\usepackage{booktabs}%
\usepackage{algorithm}%
\usepackage{algorithmicx}%
\usepackage{algpseudocode}%
\usepackage{listings}%
\usepackage[,dvipsnames]{xcolor}
\theoremstyle{thmstyleone}%
\theoremstyle{thmstyletwo}%

\theoremstyle{thmstylethree}%

\def\ga{\,\,\raise0.14em\hbox{$>$}\kern-0.76em\lower0.28em\hbox
{$\sim$}\,\,}
\def\la{\,\,\raise0.14em\hbox{$<$}\kern-0.76em\lower0.28em\hbox
{$\sim$}\,\,}

\begin{document}

\title[Article Title]{
Statistical framework for nuclear parameter uncertainties in nucleosynthesis modeling of r- and i-process
%Statistical framework for quantifying nuclear parameter uncertainties and propagating them in nucleosynthesis models
%Statistical nuclear uncertainties determination and impact on various nucleosynthesis processes
}
   % \subtitle{Impact on i-process and r-process nucleosynthesis}


\author*[1]{\fnm{Sébastien} \sur{Martinet}}\email{sebastien.martinet@ulb.be}

\author[1]{\fnm{Stephane} \sur{Goriely}}

\author[1]{\fnm{Arthur} \sur{Choplin}}

\author[1]{\fnm{Lionel} \sur{Siess}}

\affil*[1]{\orgdiv{Institut d'Astronomie et d'Astrophysique}, \orgname{Universit\'e Libre de Bruxelles (ULB)}, \orgaddress{\street{CP 226, B-1050}, \city{Brussels}, \country{Belgium}}}


\abstract{Propagating nuclear uncertainties to nucleosynthesis simulations is key to understand the impact of theoretical uncertainties on the predictions, especially for processes far from the stability region, where nuclear properties are scarcely known. While systematic (model) uncertainties have been thoroughly studied, the statistical (parameter) ones have been more rarely explored, as constraining them is more challenging. We present here a methodology to determine coherently parameter uncertainties by anchoring the theoretical uncertainties to the experimentally known nuclear properties through the use of the Backward Forward Monte Carlo method. We use this methodology for two nucleosynthesis processes: the intermediate neutron capture process (i-process) and the rapid neutron capture process (r-process). We determine coherently for the i-process the uncertainties from the (n,$\gamma$) rates while we explore the impact of nuclear mass uncertainties for the r-process. The effect of parameter uncertainties on the final nucleosynthesis is in the same order as model uncertainties, suggesting the crucial need for more experimental constraints on key nuclei of interest. We show how key nuclear properties, such as relevant (n,$\gamma$) rates impacting the i-process tracers, could enhance tremendously the prediction of stellar evolution models by experimentally constraining them.
}




\maketitle

\section{Introduction}\label{sec1}

%The Introduction section, of referenced text \cite{bib1} expands on the background of the work (some overlap with the Abstract is acceptable). The introduction should not include subheadings.

%Springer Nature does not impose a strict layout as standard however authors are advised to check the individual requirements for the journal they are planning to submit to as there may be journal-level preferences. When preparing your text please also be aware that some stylistic choices are not supported in full text XML (publication version), including coloured font. These will not be replicated in the typeset article if it is accepted. 

In nuclear astrophysics, models that predict nuclear reaction rates, nuclear masses, and other fundamental properties often rely on numerous adjustable parameters. Parameter uncertainties stem from the incomplete experimental knowledge or theoretical understanding of these values within a model. Unlike model uncertainties, which originate from the intrinsic limitations of the physical assumptions or approximations in a model, parameter uncertainties reflect the model defects and the variability in parameter choices that still allow the model to reproduce known experimental data. For example, nuclear models such as the Hartree-Fock-Bogolyubov (HFB) model \citep{Goriely13a} are optimized against measured data, but small local variations in these parameters can yield different outcomes when extrapolated to untested regions.

Determining parameter uncertainties in nuclear astrophysics is particularly challenging due to the complex, multidimensional nature of nuclear models and the vast number of adjustable parameters they require \citep{Goriely14}. Each nuclear model is finely tuned against experimental data, yet no model is completely parameter-free. Because these parameters are often based on extrapolations from limited experimental data, even small variations can lead to significant discrepancies when predicting properties of nuclei far from stability. Parameter uncertainties, therefore, reflect a critical but complex dimension of model reliability, especially when applied to nuclei that cannot be directly measured.

Historically, attempts to account for these uncertainties were often oversimplified. A common approach involved applying arbitrary scaling factors to nuclear rates or masses, multiplying them by fixed values to simulate uncertainty ranges \citep{Mumpower16, Surman16, Nikas20, Jiang21}. Although convenient, this method fails to capture the actual complexity of parameter variations and their correlated effects across different nuclear properties. Such an arbitrary scaling ignores the fact that uncertainties are rarely uniform across all parameters and do not adequately reflect the nuanced interactions between different model parameters. Consequently, these simplistic methods can lead to either overestimated or underestimated uncertainties, especially when applied to sensitive astrophysical scenarios like the r- and i-process nucleosynthesis in neutron-rich environments \citep{Martinet24}.

In this work, we present a coherent approach to determining the parameter nuclear uncertainties and the subsequent impact of these uncertainties on the prediction of nucleosynthesis for different processes thought to occur in astrophysics environment. Section \ref{Sect:Method} presents a statistical approach to the parameter uncertainty determination known as the Backward-Forward Monte Carlo (BFMC) method. Section \ref{Sect:i-pro} describes the determination of neutron capture rates parameter uncertainties and their impact on the i-process nucleosynthesis in early Asymptotic Giant Branch (AGB) stars. Section \ref{Sect:r-pro} explore the more complex determination of nuclear mass uncertainties and its impact on the r-process nucleosynthesis in neutron star mergers (NSM). Finally, conclusions and perspectives are presented in Section \ref{Sect:conclusion}. 

\section{Method}
\label{Sect:Method}
Accurately quantifying and propagating parameter uncertainties in nuclear models requires a robust methodological approach. The Backward-Forward Monte Carlo (BFMC) method provides a systematic framework for constraining model parameters based on experimental data and propagating these uncertainties to model predictions.
\subsection{Backward-Forward Monte Carlo (BFMC) methodology}
To accurately quantify parameter uncertainties and propagate them to model predictions, the BFMC approach provides a structured method that integrates statistical sampling with experimental constraints. The BFMC method, introduced in nuclear data evaluations to address parameter sensitivity \cite{Chadwick07,Bauge11}, consists of two main steps. 

In the backward step, the model parameters are sampled within a range informed by experimental constraints, and each parameter set is evaluated against these constraints using a $\chi^2$ estimator or similar criterion. This selection process identifies only the parameter sets that accurately reproduce experimental data, thereby limiting the parameter space to realistic variations. The backward step thus restricts the model to a “constrained” set of parameters, ensuring that only physically meaningful variations are included in the subsequent analysis.

The forward step uses these constrained parameter sets to simulate the desired quantities for conditions where experimental data may be sparse or unavailable, such as reaction rates for neutron-rich nuclei or masses far from stability. By running numerous forward simulations with these varied parameter sets, the BFMC approach propagates the parameter uncertainties through the model, yielding a distribution of predicted outcomes. This approach enables us to quantify the sensitivity of model outputs to parameter variability in a coherent and systematic manner, capturing the potential spread in predictions due to uncertainties in underlying parameters.

The BFMC method is particularly advantageous for applications in nuclear astrophysics, where understanding the uncertainties in predicted abundances or reaction rates can influence our interpretation of nucleosynthesis pathways in astrophysical environments. For example, in modeling the r-process within neutron star mergers, the BFMC method enables an accurate assessment of the range of possible nucleosynthetic yields based on known mass uncertainties, helping to refine predictions for the elemental compositions observed in kilonova ejecta. Similarly, for the i-process in low-metallicity stars, the BFMC method allows for a robust estimate of how nuclear parameter uncertainties impact predicted surface abundances, informing observational comparisons.

\subsection{Handling and applying BFMC-derived parameter uncertainties}
\label{sect:applying_BFMC}
% The parameter uncertainties obtained from the BFMC approach are uncorrelated. This means that any rates in-between the maximum and the minimum values obtained from the BFMC can be used in a nuclear network. 
% However, this is only acceptable in a nuclear network if it is used within the same nuclear model. Otherwise, the correlations would be neglected and would lead to an overestimation of the uncertainties.

% Sensitivity studies using the MC method should be applied within the maximum and minimum limits obtained by the BFMC. %using the same model across the same nuclear network. 
% Using these values to derive a deviation or the ratio between max/min and then applying it to another nominal rate value would neglect the correlations between them. We recommend using these values directly as limits to the Markov Chain Monte Carlo (MCMC) exploration method. Note that these values are obtained as maximum and minimum rates coherent with experimental constraints, but nothing leads to thinking that they are the limits of a normal distribution centered on nominal rates. We would advise considering a uniform distribution between these limits to apply the MC method. 

The parameter uncertainties obtained from the BFMC approach are uncorrelated. This means that any rates between the maximum and minimum values determined by the BFMC can be used in a nuclear network. However, this is only valid if the same nuclear model is consistently used; applying these values across different models would neglect correlations and lead to an overestimation of uncertainties.

Sensitivity studies using the MC method should remain within the maximum and minimum limits provided by the BFMC. Using these limits to calculate deviations or max/min ratios and applying them to another nominal rate value disregards the inherent correlations between parameters. We recommend using these BFMC-derived limits directly as limits to the Markov Chain Monte Carlo (MCMC) exploration method. It is important to note that while these values represent coherent maximum and minimum rates based on experimental constraints, they are not necessarily the limits of a normal distribution centered on nominal rates. Instead, a uniform distribution between these limits should be considered when applying the MC method.

\section{Determining coherently parameter uncertainties for neutron capture rates: application to the i-process in AGB stars}
\label{Sect:i-pro}

The intermediate neutron capture process, or i-process, introduced by \citet{cowan77}, is a neutron capture process occurring in various astrophysical environments \citep[see][]{choplin21}, notably in low-metallicity, low-mass AGB stars \citep[e.g.,][]{iwamoto04, cristallo09b, suda10, choplin21, choplin22, goriely21, gilpons22}. The i-process is triggered by a proton ingestion event (PIE) where protons are mixed into a convective helium-burning zone. In AGB stars, PIEs can arise during the early thermally pulsing (TP) phase, when the convective thermal pulse overcomes the entropy barrier at the base of the H-burning shell, and engulfs protons. These protons are transported downward, burning via $^{12}$C($p,\gamma$)$^{13}$N and decaying to $^{13}$C, which initiates the $^{13}$C($\alpha,n$)$^{16}$O reaction at temperatures near 250 MK. This reaction raises neutron densities of $\sim 10^{15}$ cm$^{-3}$, allowing i-process nucleosynthesis to proceed. The large energy released by the $^{12}$C($p,\gamma$)$^{13}$N reaction splits the convective pulse \citep[see][]{choplin22} and the upper part of the TP ends up merging with the envelope, where the i-process elements are mixed and eventually expelled by stellar winds.

\begin{figure}
    \centering
    \includegraphics[width=0.8\linewidth]{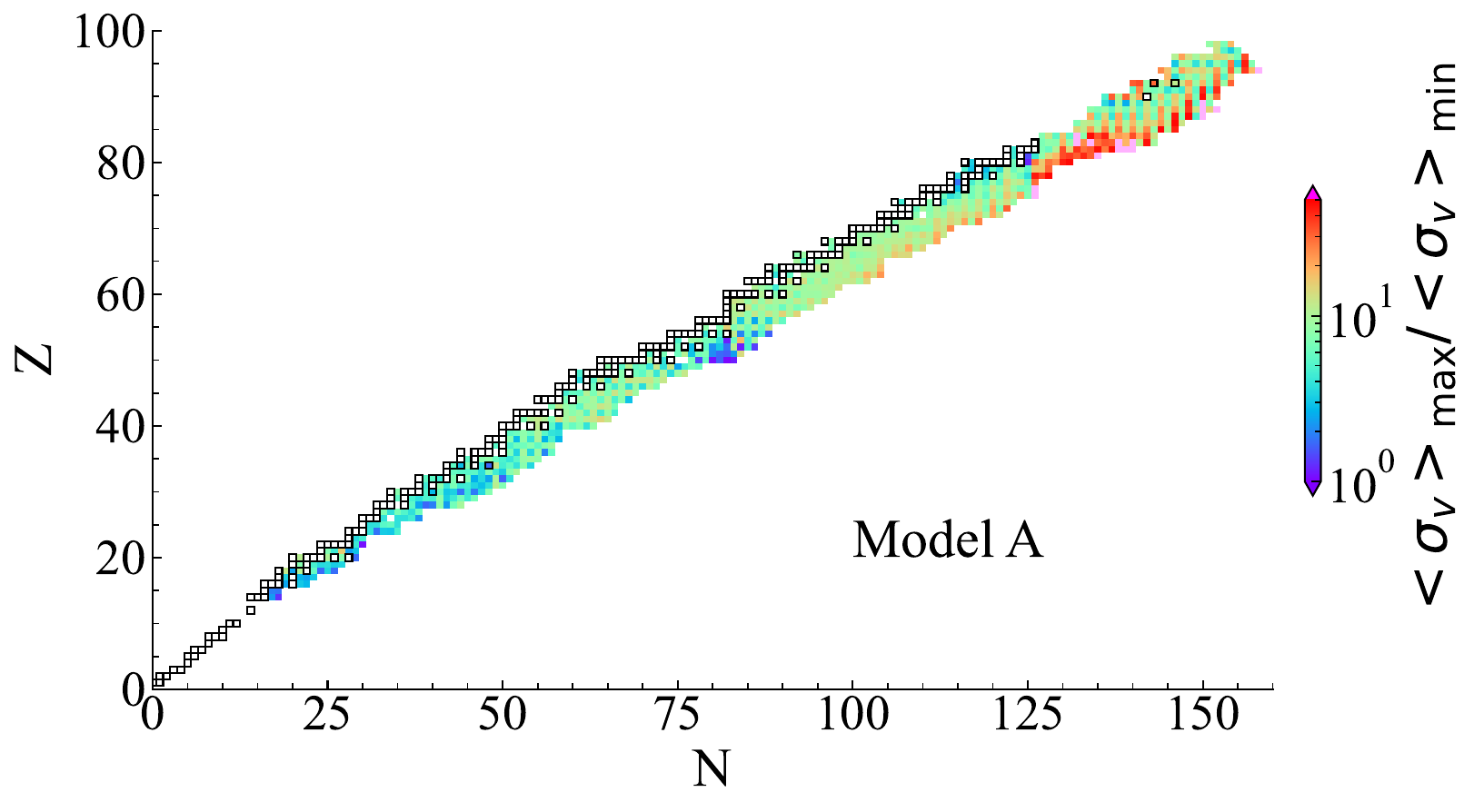} 
    \includegraphics[width=0.8\linewidth]{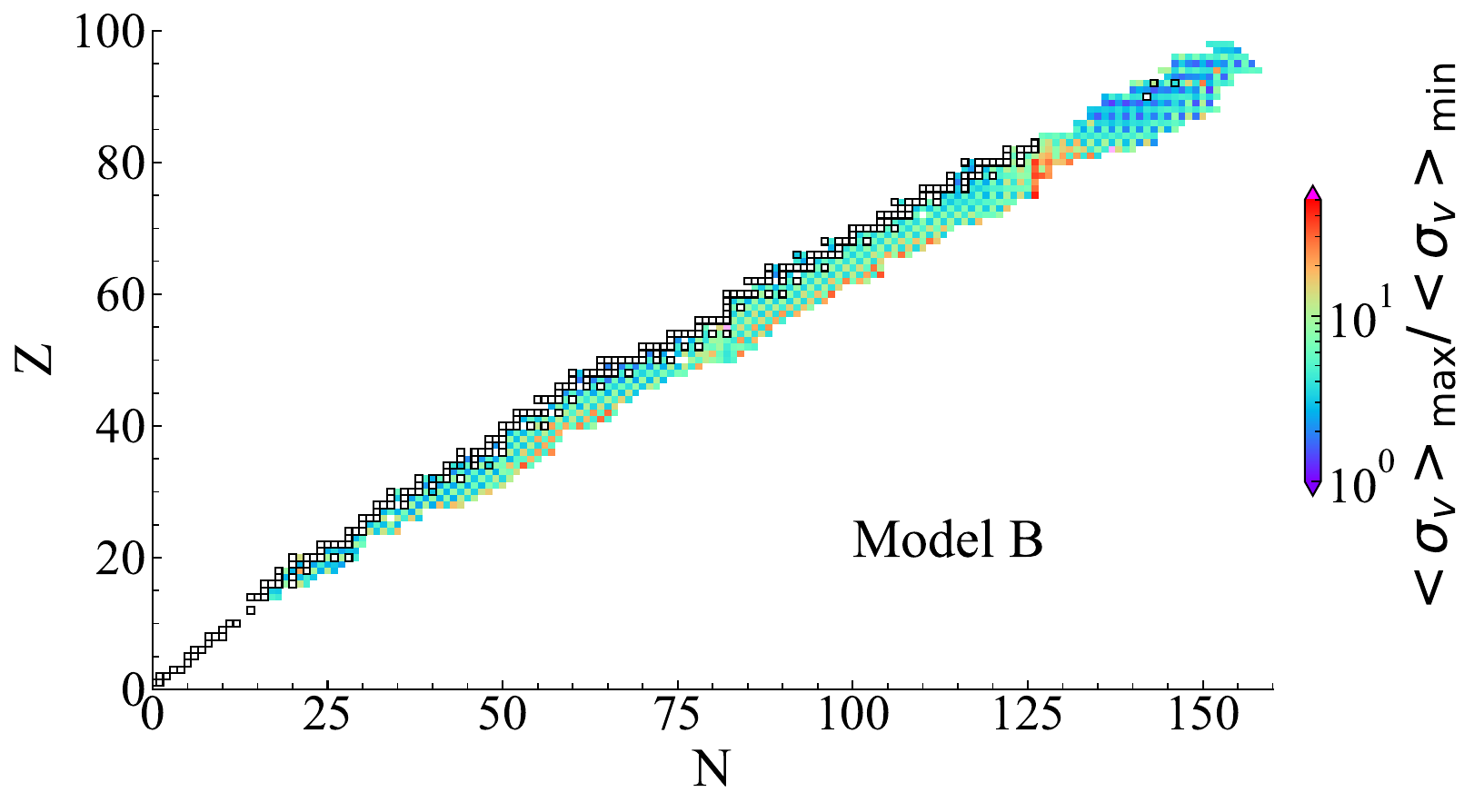}
    \caption{Parameter uncertainties affecting neutron capture rates using Model A (upper panel) or B (lower panel) for an i-process nuclear network.  See text for more details. }
    \label{fig:param_uncertainty_ipro_rates}
\end{figure}

The nucleosynthesis occurring in this process is highly dependent on the neutron capture rates. Although important efforts have been made to measure neutron capture rates at astrophysical energies, only a limited number of reactions have been constrained outside the stability region. Only $\sim$240 neutron capture rates are experimentally known, while our i-process nuclear network is made up of more than 1000 nuclei. This means that a large part of the nucleosynthesis is sensitive to theoretical nuclear uncertainties.

\subsection{Neutron capture rates and their parameter uncertainties}

% \subsection{Nuclear Uncertaintes impai }
We assess parameter uncertainties for two nuclear models characterized by two different combinations of nuclear level density (NLD) and photon strength function (PSF) models for the calculation of the neutron capture rates with the TALYS reaction code \cite{Koning23}. The first set, A, adopts the HFB plus combinatorial NLDs \citep{Goriely08b} and the D1M+QRPA PSFs for both the electric E1 and magnetic M1 dipole components \citep{Goriely18a}. While set A is based on rather microscopic ingredients, set B considers more phenomenological models, namely the constant-temperature model of NLDs \citep{Koning08} and the Lorentzian-type SMLO model for the PSFs \citep{Goriely18b}.

We use the BFMC method to estimate the uncertainties that arise from our four model parameters. We used as $\chi^2$ estimator the $f_{\rm rms}$ indicator with respect to the 239 experimental (n,$\gamma$) MACS at 30 keV from the KADoNiS database \citep{Dillmann06}. This backward step allows us to pursue our theoretical computations with constrained combinations of parameters for the NLD and PSF. See Ref.~\cite{Martinet24} for more details.

% In Fig.~\ref{fig:param_uncertainty_ipro_rates}, the resulting uncertainties obtained for the neutron capture rates involved in the i-process are shown for both nuclear models. They are color-coded showing the ratio between the maximum and minimum ratios obtained through BFMC parameter uncertainties. The uncertainties are smaller close to the stability region due to the coherent BFMC propagation of the experimental uncertainties. When straying further away to neutron-rich nuclei, the neutron capture rates can lead to uncertainties up to a factor of 100. A striking feature is the much reduced uncertainties obtained around the actinides for Model B compared to Model A. This is due to the fact that the NLDs in Model B are computed with the constant-temperature formula until a certain matching temperature $E_m$ where it switches to a Fermi gas model \cite{Koning08}. As we only vary the constant-temperature parameters $E_0$ and $T$ in the BFMC method and not the Fermi gas parameters, NLDs are essentially affected at excitation energies ($U<E_m$). However, for actinides, $E_m$ is mainly lower than the neutron separation energy, inducing an underestimated impact on the uncertainties in Model B. 

The resulting parameter uncertainties for neutron capture rates are shown in Fig.~\ref{fig:param_uncertainty_ipro_rates} for both nuclear models, with color coding indicating the ratio between the maximum and minimum rates obtained through BFMC. Uncertainties are lower near the stability region due to the coherent propagation of experimental uncertainties in the BFMC method, but they can increase up to a factor of 100 for neutron-rich nuclei. Notably, uncertainties around the actinides are significantly lower in Model B compared to Model A. This difference arises because Model B computes nuclear level densities (NLDs) using a constant-temperature formula up to a matching energy $E_m$, where it transitions to a Fermi gas model \cite{Koning08}. Since we only vary the constant-temperature parameters $E_0$ and $T$ in the BFMC method, the NLDs in Model B are primarily affected at excitation energies below $E_m$. For actinides, $E_m$ is often lower than the neutron separation energy, leading to an underestimation of uncertainties in Model B.

\begin{figure}
    \centering
    \includegraphics[width=0.98\linewidth]{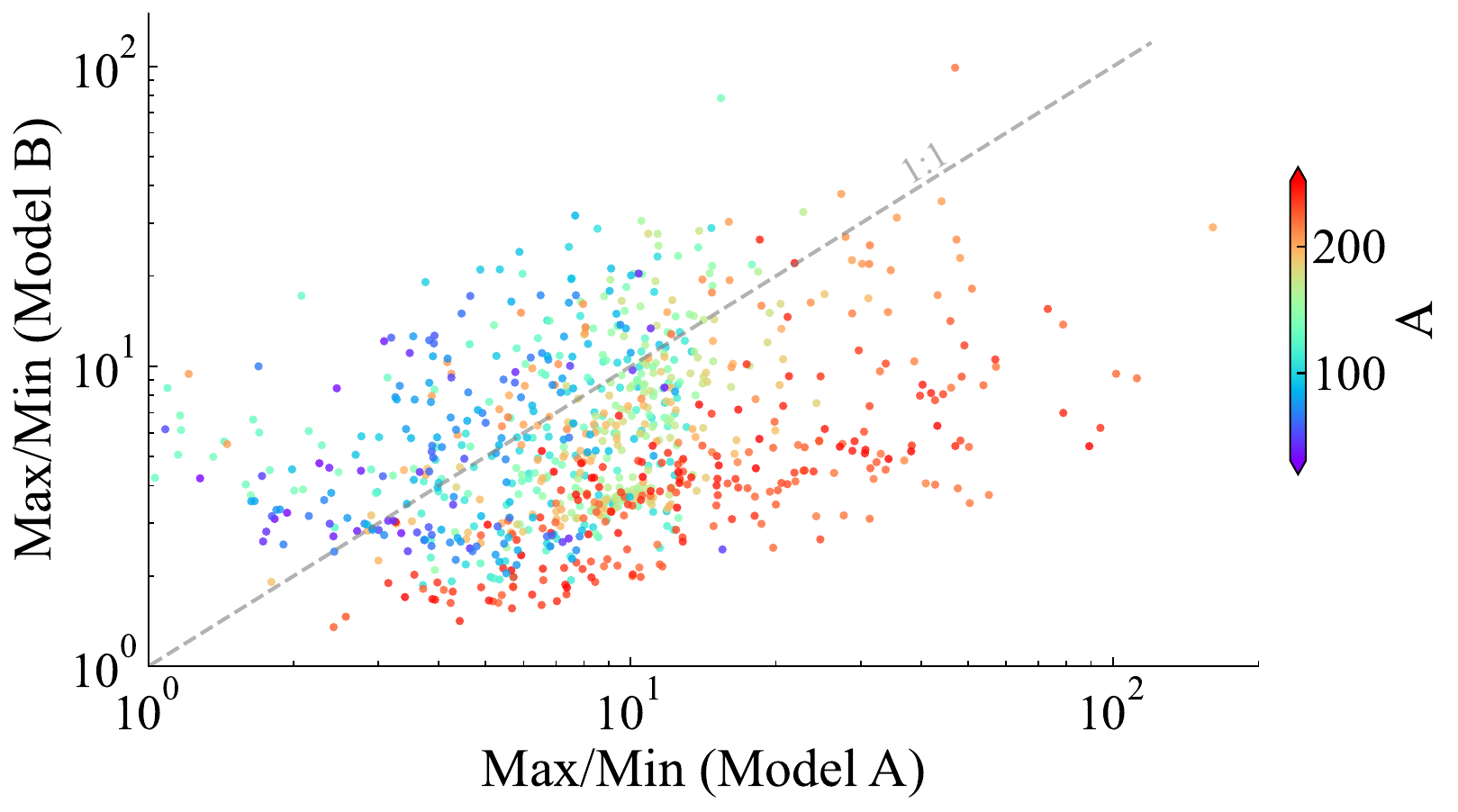} 
    \caption{Illustration of the parameter uncertainties on neutron capture rates obtained by their maximum-to-minimum ratios with Model A vs those of Model B. Each point is color-coded by the nuclear mass number $A$ of the target nucleus. }
    \label{fig:max_min_model_A_vs_B}
\end{figure}

Figure \ref{fig:max_min_model_A_vs_B} illustrates the impact of the parameter uncertainties of both models on neutron capture rates. The ratio of maximum to minimum rates for Model B is plotted against those of Model A, where each point is color-coded by the nuclear mass number $A$ of the target nucleus. The diagram reveals no clear correlation between the uncertainties of the two models, as indicated by the large dispersion of points from the 1:1 relationship. 
For nuclei with $A>200$, Model A consistently shows much larger uncertainties. As shown in Fig.~\ref{fig:param_uncertainty_ipro_rates}, this is primarily due, as explained above, to Model B switching regime which reduces the influence of parameter variations used in this study for neutron capture rates by actinides.

The absence of correlation in Fig. \ref{fig:max_min_model_A_vs_B} shows that different nuclear models do not lead to the same uncertainty ratio. Hence, those uncertainty ratios are correlated to the model and should only and coherently be used for this model alone. Using these ratios to estimate  uncertainties upon nominal rates obtained with another nuclear model would fail to include the correct correlations, hence lead to over- or under-estimation of the uncertainties. As discussed in Sect. \ref{sect:applying_BFMC}, we advise not to use the maximum-to-minimum ratios obtained from BFMC, but directly the maximum and minimum rates as limits to sensitivity studies.

\begin{figure}
    \centering
    \includegraphics[width=0.99\linewidth]{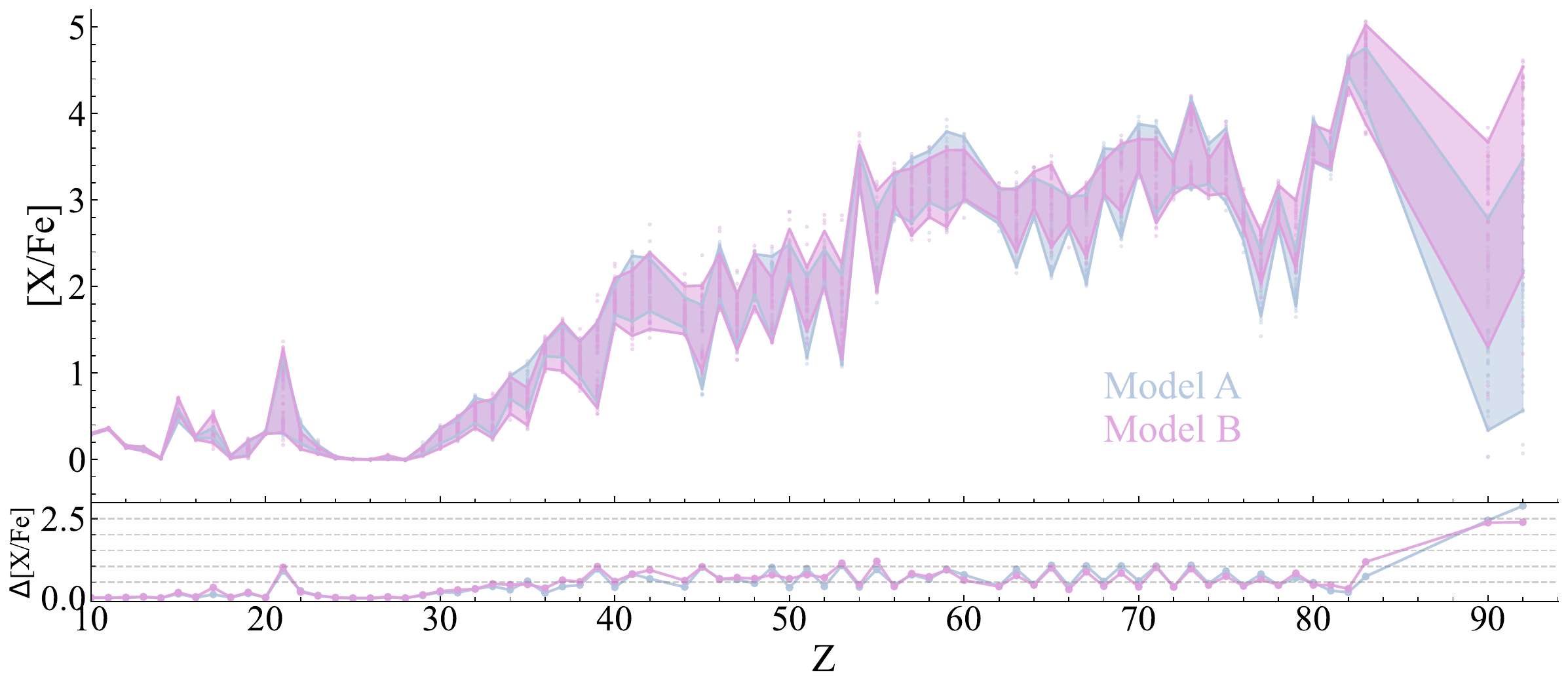}
    \caption{Impact of the neutron capture rates parameter uncertainties on the i-process nucleosynthesis during the early AGB phase of a low-mass low-metallicity star.}
    \label{fig:param_uncertainty_ipro}
\end{figure}

\subsection{Impact of the neutron capture rate uncertainties on the i-process}

To propagate parameter uncertainties to nucleosynthesis simulations, we generated 50 sets of rates, each containing randomly selected minimum or maximum values for the 868 unknown (n,$\gamma$) reaction rates. These values are based on the maximum and minimum rates obtained using the BFMC method. Due to the uncorrelated nature of these uncertainties, rates can be chosen at random within this range. The resulting random sets are then used in stellar evolution models to evaluate the impact of parameter uncertainties on the final surface abundances of our 1$M_{\odot}$, [Fe/H]~$=-2.5$ AGB star. The resulting uncertainties obtained for both nuclear models are shown in Fig.~\ref{fig:param_uncertainty_ipro}. We retained only final surface abundances within the 5th to 95th percentiles to account for numerical uncertainties. Odd-$Z$ elements are affected by a 1 dex uncertainty on average and the even-$Z$ elements by a 0.5 dex uncertainty. This feature comes from odd-$Z$ elements having mostly one stable isotope, hence being more sensitive to the uncertainty of one single reaction rate. The parameter uncertainties are relatively similar between Models A and B, underlying the fact that the lack of experimental constraints is the main source of uncertainty for the i-process nucleosynthesis. The actinide production is relatively sensitive to both model and parameter uncertainties due to the involvement of key reactions far from the stability region enabling the nuclear flow to by-pass the $\alpha$-decaying nuclei heavier than Pb \cite{Choplin2022letter}, as further discussed below.

\begin{figure}
    \centering
    \includegraphics[width=0.99\linewidth]{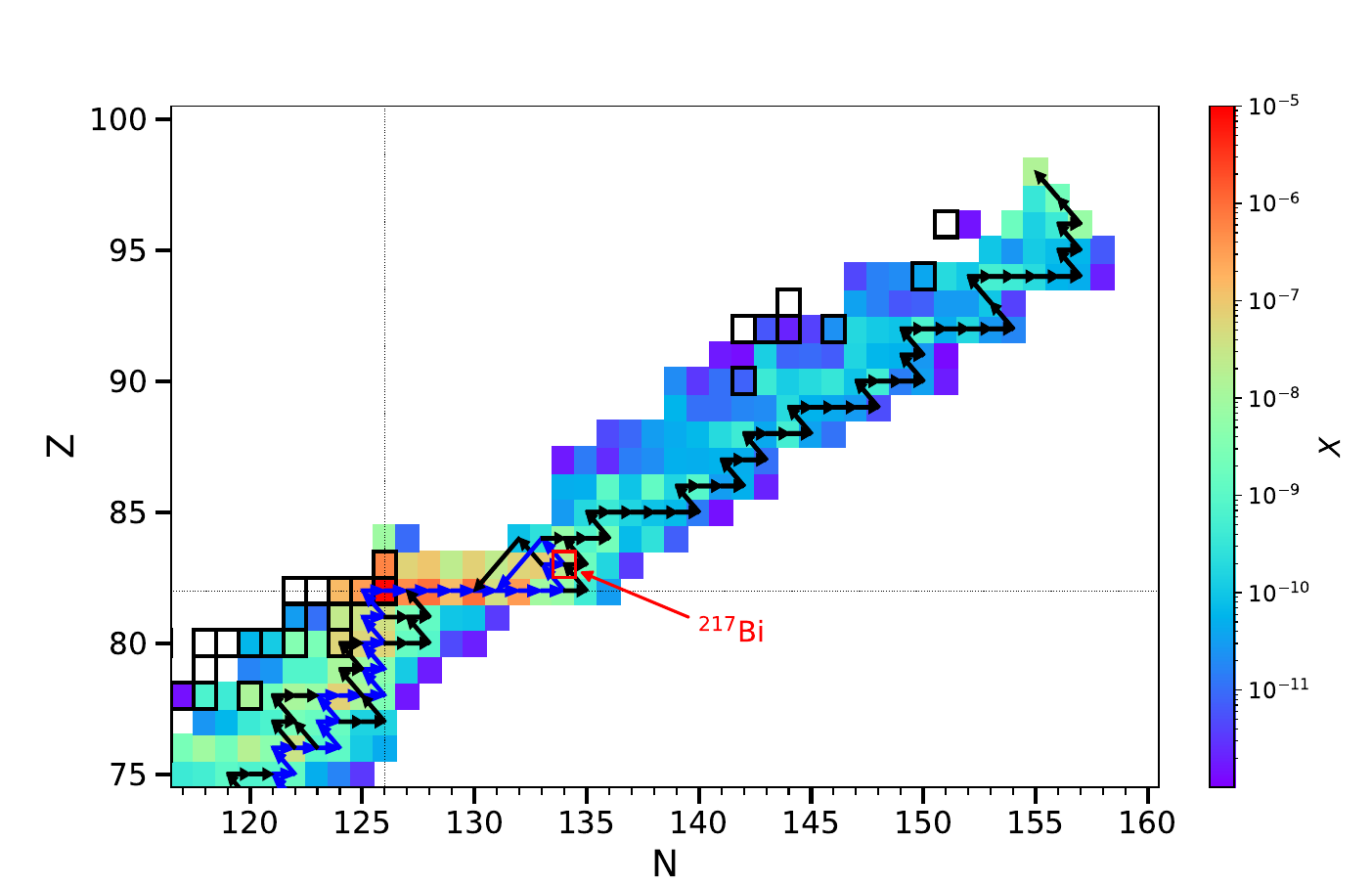}
    \caption{Primary (blue) and secondary (black) i-process paths (starting from $^{56}$Fe) in a 1~$M_{\odot}$ AGB model at [Fe/H]~$=-2.5$, taken at the base of the convective TP during peak neutron density. A path is marked as secondary if it carries at least 30\% of the total flux. The color of each nucleus gives its mass fraction at this time.}
    \label{fig:Flux_i_process}
\end{figure}

\begin{figure}
    \centering
    \includegraphics[width=0.99\linewidth]{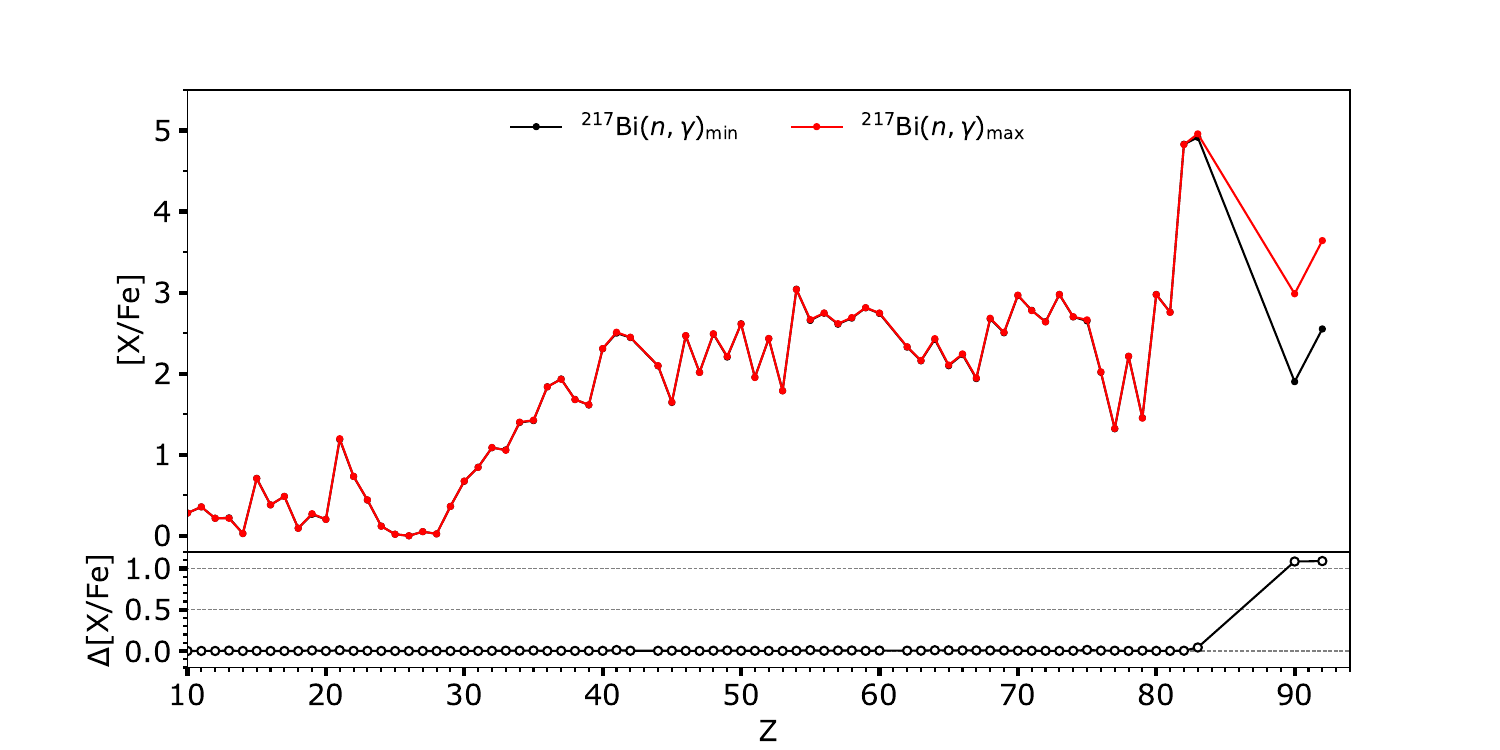}
    \caption{Impact of the $^{217}$Bi(n,$\gamma$)$^{218}$Bi rate parameter uncertainty on the i-process nucleosynthesis during the early AGB phase of a low mass low metallicity star.}
    \label{fig:bi217_uncertainty}
\end{figure}

Using statistical methods, important key reactions can be identified \citep{Martinet24} and their impact on abundances estimated. We focus here on the $^{217}$Bi(n,$\gamma$)$^{218}$Bi, a reaction that turns out to have a significant impact on the production of actinides. Fig. \ref{fig:Flux_i_process} shows the i-process path in our AGB models, highlighting the importance of reactions around the Pb-Bi-Po isotopes, acting as a gate to the secondary path for producing actinides. %The $^{217}$Bi(n,$\gamma$) is highlighted and pre

Figure \ref{fig:bi217_uncertainty} shows the impact of the $^{217}$Bi(n,$\gamma$)$^{218}$Bi parameter uncertainty on the i-process nucleosynthesis. Two stellar models are computed with maximum and minimum $^{217}$Bi(n,$\gamma$) rates, all the other rates being untouched (and equal to the geometrical mean between their minimum and maximum values).    
As expected from the statistical analysis, this rate has a direct impact on actinide production, with more than 1 dex on the total uncertainty. Constraining this rate would then greatly reduce the uncertainties on the potential production of actinide by the i-process. Note that the total nuclear uncertainty on actinides presented in Fig. \ref{fig:param_uncertainty_ipro} shows a total uncertainty of around 2 dex (for U and Th). This shows how the complex interplay between different max/min rates impacts the uncertainties. Resolving the $^{217}$Bi(n,$\gamma$)$^{218}$Bi uncertainty would not reduce fully the uncertainties on the actinide production, but as this reaction acts as a gate to the secondary flux above the Pb region, it would allow determining if the flux is strong enough to give rise to a noticeable overproduction of Th and U.

\section{Determining coherently parameter uncertainties for nuclear masses: application to r-process in NSM}
\label{Sect:r-pro}

The r-process in stellar nucleosynthesis is crucial for explaining the formation of stable and long-lived radioactive neutron-rich nuclides heavier than iron observed across stars of various metallicities, including in the Solar System \citep[see][]{Arnould07, Arnould20, Cowan21}.
Advanced simulations support NSM as an effective r-process site, producing elements up to the third abundance peak and the actinides. This enrichment arises from both the prompt material expelled during the dynamical phase and the outflows from the post-merger evolution of the neutron star (NS)-torus and black-hole (BH)-torus systems \citep{Just15, Just23}. The resulting abundance distributions align well with those observed in the Solar System and in low-metallicity stars \citep{Cowan21}.

However, uncertainties in r-process modeling persist, especially concerning the extended nuclear physics inputs \citep[e.g.,][]{Mendoza15, Kullmann23}. A significant source of uncertainty remains the unknown nuclear masses of extremely exotic neutron-rich nuclei produced by the r-process under NSM conditions. High neutron densities and temperatures are responsible for the competition between radiative neutron captures and photoneutron emissions within isotopic chains. This competition is essentially governed by the neutron separation energy ($S_n$), hence by nuclear masses. Yet predicting the impact of mass uncertainties on the ejecta composition is challenging. This difficulty arises from the neutron separation energy which embeds a correlated mass difference. Similarly, the nuclear flow towards heavier elements is controlled by $\beta$-decays, hence by the $\beta$-decay ($Q_{\beta}$) energy also embeds a correlated mass difference. Propagating mass uncertainties into $S_n$ and $Q_\beta$, as well as into reaction and decay rates and finally nucleosynthesis observables remains consequently a difficult task that is far from being solved at present \citep[see e.g.,][]{Goriely15c, Mendoza15, Sprouse20, Mumpower16, Kullmann23}.

We present here an improved description of the parameter uncertainties affecting nuclear masses and their propagation into reaction rates and r-process abundances in NSM. 

\subsection{Nuclear masses and their parameter uncertainties}
Parameter uncertainties in HFB-24 masses were evaluated using the BFMC method \citep{Goriely14}. At the end of the backward MC step, a subset of $N^\xi_{\rm comb} = 10,931$ parameter combinations, constrained by experimental masses ($\sigma_{M,{\rm exp}} \le 0.8$~MeV), was selected. This subset was then used in the forward MC step to calculate approximately 5'000 unknown masses of neutron-rich nuclei. This method yields uncorrelated uncertainties by design. Notably, \citet{Goriely14} assumed that local parameter changes do not impact deformation energies. While larger uncertainties may apply to highly deformed nuclei, the results remain accurate for spherical nuclei, especially those near neutron shell closures critical for r-process applications.

The separation energy $S_n$ of a ($Z$,$N$) nucleus is crucial during r-process neutron irradiation, as it determines the main nuclei produced within an isotopic chain through the $(n,\gamma) \rightleftharpoons (\gamma,n)$ competition \citep{Goriely92}. It is defined as
\begin{equation}
    S_n(Z,N)=M(Z,N-1)+m_{\rm neut}-M(Z,N) \quad ,
     \label{eq:Sn}
\end{equation}
where $M$ is the atomic mass and $m_{\rm neut}$ the neutron mass.
Similarly $Q_\beta$, defined as
\begin{equation}
    Q_\beta(Z,N)=M(Z,N)-M(Z+1,N-1) \quad ,
    \label{eq:Qb}
\end{equation}
is an important quantity affecting the energy production, but also the $\beta$-decay rates, hence the composition of the ejecta. 

To estimate the parameter uncertainties affecting the neutron separation energy $S_n$, we consider two approaches. In the first one, the BFMC method is used to determine mass uncertainties $\sigma_M$ for all unknown nuclei, which are then propagated to $S_n$ using Eq.~\eqref{eq:Sn}. The second approach applies the BFMC method directly to $S_n$, calculating both masses that define $S_n$ coherently within the same parameter set selected by BFMC.

\begin{figure}
    \centering
    \includegraphics[width=0.99\linewidth]{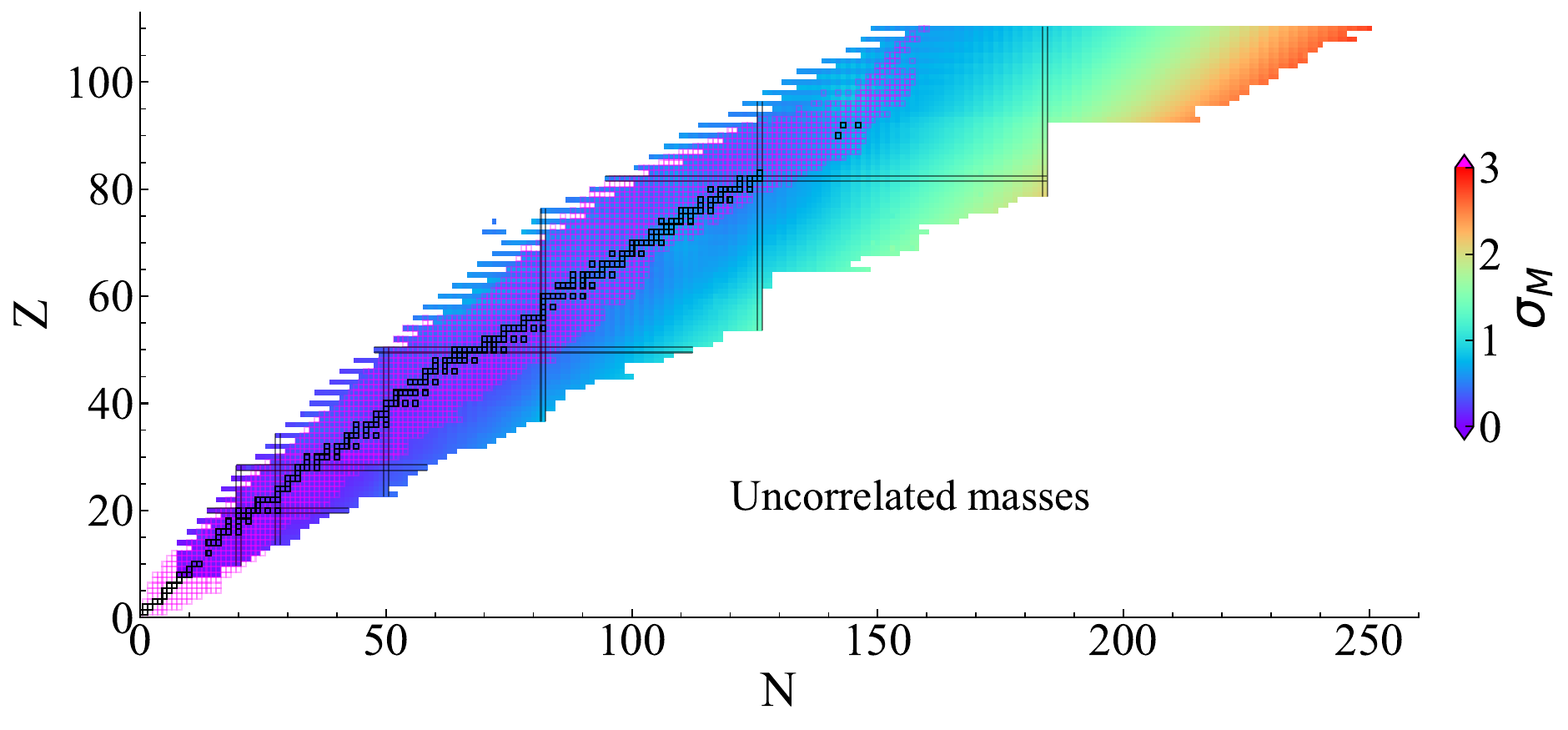}
    \includegraphics[width=0.99\linewidth]{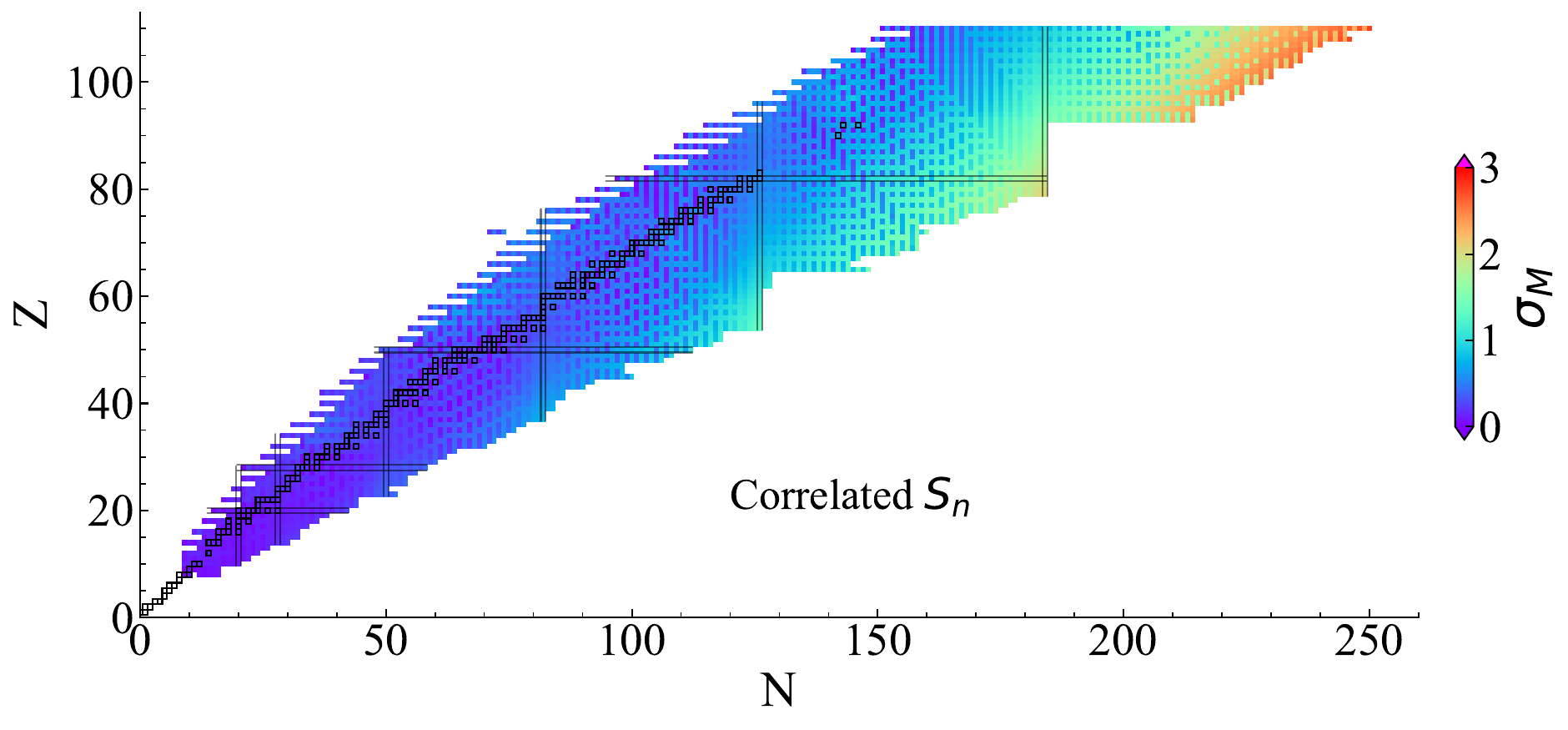}
    \caption{Top panel: Representation in the ($N$,$Z$) plane of the uncorrelated masses uncertainties $\sigma_M(Z,N)$ (in MeV) obtained from 10931 runs with $\protect\sigma_{M,{\rm exp}} <0.8$~MeV. See \citet{Goriely14} for more details. The nuclei outlined with magenta open squares are the 2550 experimentally known masses \cite{Wang21}. {The shells closures are displayed by solid lines.} Bottom panel: same for the case where we maximize/minimize $S_n$ while keeping correlations and coherent masses. The experimentally known masses are not highlighted for more clarity.}
    \label{fig:masses_uncertainties}
\end{figure}

Figure \ref{fig:masses_uncertainties} shows the masses uncertainties for both cases. The top panel shows the uncorrelated masses uncertainties. 
%The nuclei highlighted in magenta represent approximately the 2,500 experimentally known masses. 
Based on the BFMC method, we anticipate that uncertainties for these nuclei will be around the rms deviation of HFB-24 relative to experimental masses, i.e., 0.55 MeV. Consequently, the nearby unknown masses are expected to have uncertainties of a similar magnitude. Parameter combinations consistent with experimental uncertainties are unlikely to cause significant variations for neighboring nuclei but tend to increase further from experimentally known regions. This trend is evident, as the largest uncertainties, up to $\sigma_M \approx 3$ MeV, are observed for neutron-rich nuclei near the neutron drip line. These uncorrelated mass uncertainties $\sigma_M$ can then be used to estimate corresponding neutron separation energies (Eq.~\ref{eq:Sn}) with their uncertainties, ignoring the correlated nature embedded in Eq.~\eqref{eq:Sn}. 

The bottom panel shows the same figure for the second case, for which we are looking to maximize the uncertainties on the $S_n$ themselves while taking into account $S_n$/$Q_\beta$ correlations embedded in Eqs.~\eqref{eq:Sn}-\eqref{eq:Qb} and at the same time ensure that a coherent set of masses is used \cite{Martinet24}. While globally the trends observed for the first case are similar, we can observe along isotopic and isobaric chains alternate low and high $\sigma_m$. The isotopic chain effect is due to the correlations imposed by the definition of the separation energy (Eq. \ref{eq:Sn}) where two neighboring masses are linked. Maximizing the $S_n$ uncertainties results in fact in picking alternating masses that are not maximizing the mass uncertainties because of $S_n$ being the difference between two neighboring masses.
The second feature we can see along isobaric chains comes from the definition of $Q_\beta$ itself (Eq. \ref{eq:Qb}) where isobaric neighbors are linked by masses. %When taking these two correlations into account, a more complex picture of mass uncertainties designed to maximize or minimize $S_n$ is found. \SG{SG: pas sur que c'est ce que tu voulais dire !}

\subsection{Impact of the nuclear mass uncertainties on the r-process}

\begin{figure}
    \centering
    \includegraphics[width=0.99\linewidth]{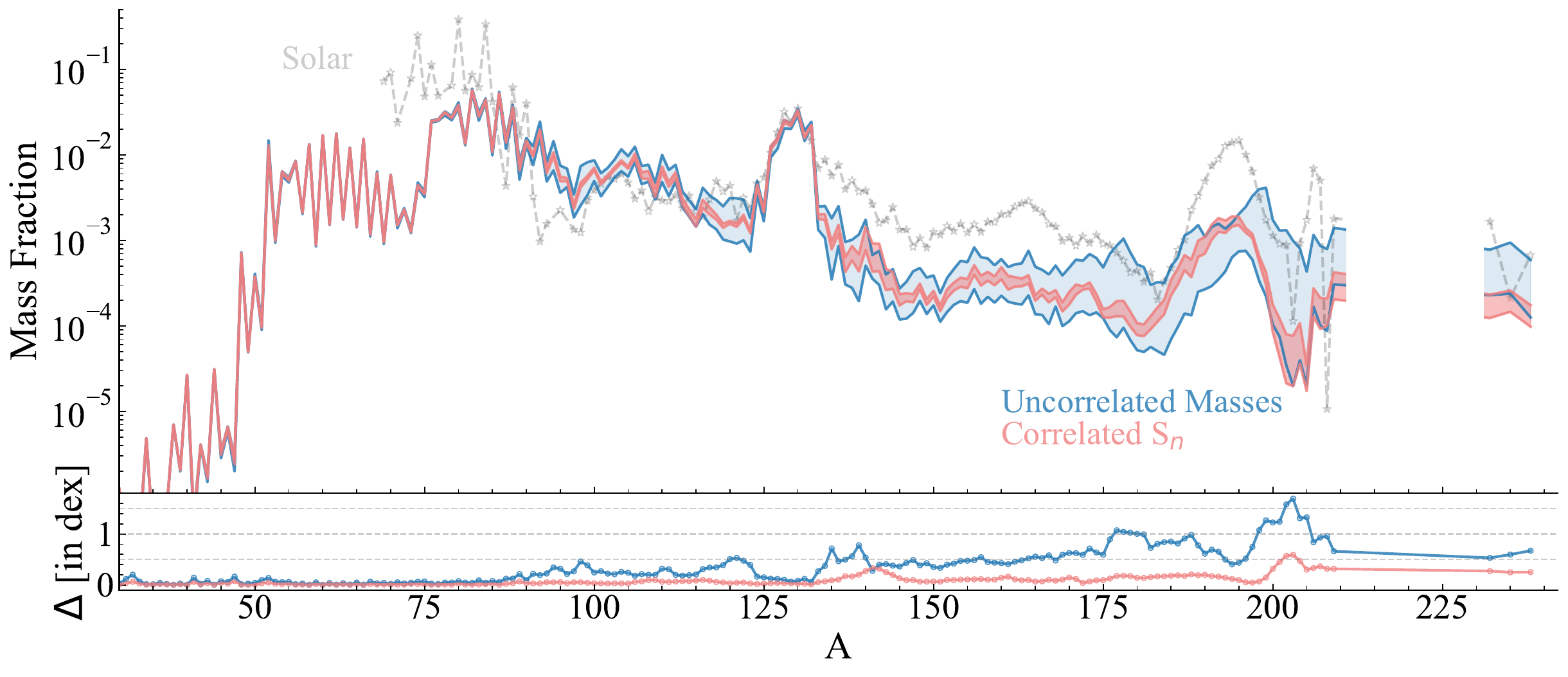}
    \caption{Impact of the nuclear mass uncertainties on the mass fractions of stable nuclei (and long-lived Th and U) of the material ejected from the binary 1.375-1.375 M$_\odot$ NSM model \cite{Just23} as a function of the atomic mass A. The blue results correspond to uncorrelated $S_n$ uncertainties and the red one to correlated $S_n$ uncertainties. Solar r-abundances (Goriely 1999) are shown in grey as a reference and are scaled to the $A = 130$ abundance in the simulation. The lower panel shows the range of the abundance uncertainties (in log scale).}
    \label{fig:case1_vs_4}
\end{figure}

Figure \ref{fig:case1_vs_4} shows the impact of the nuclear uncertainties on the r-process nucleosynthesis in a NSM for both cases presented above, i.e. considering uncorrelated or correlated $S_n$ estimated from HFB-24 parameter uncertainties. The lower panel shows the range of the abundance uncertainties. 
The solar r-abundances, shown in grey for reference, are scaled to match the $A=130$ abundance from the  simulation using uncorrelated masses. Since an (n,$\gamma$)--($\gamma$,n) equilibrium is established in most ejecta trajectories, nuclear masses play a critical role in determining where this equilibrium occurs within each isotopic chain. Propagating larger $S_n$ uncertainties for neutron-rich nuclei results in larger  changes in r-process yields.

For elements up to Ni, nucleosynthesis primarily occurs in trajectories with a relatively high electron fraction ($Y_e \ga 0.4$) \citep{Just23}, minimizing the impact of mass uncertainties on nuclei with $A \la 70$. In contrast, heavier elements are produced in low-$Y_e$ trajectories, where exotic neutron-rich nuclei with significant mass uncertainties dominate. This is particularly evident for mass fractions with $A > 140$, where these uncorrelated mass uncertainties lead to large deviations in the predicted composition of the ejecta. These deviations are notably reduced when adopting uncertainties from correlated $S_n$.
It is worth emphasizing that this second case is not merely a subset of the first one. As a result, some abundances observed in the correlated $S_n$ case, such as those around $A \simeq 210$, are absent in the uncorrelated masses case. Additionally, the overestimated uncertainties of the latter, including the potential weakening of the $N=126$ shell closure, lead to a third r-process peak that is either flattened or diminished.

\begin{figure}
    \centering
    \includegraphics[width=0.99\linewidth]{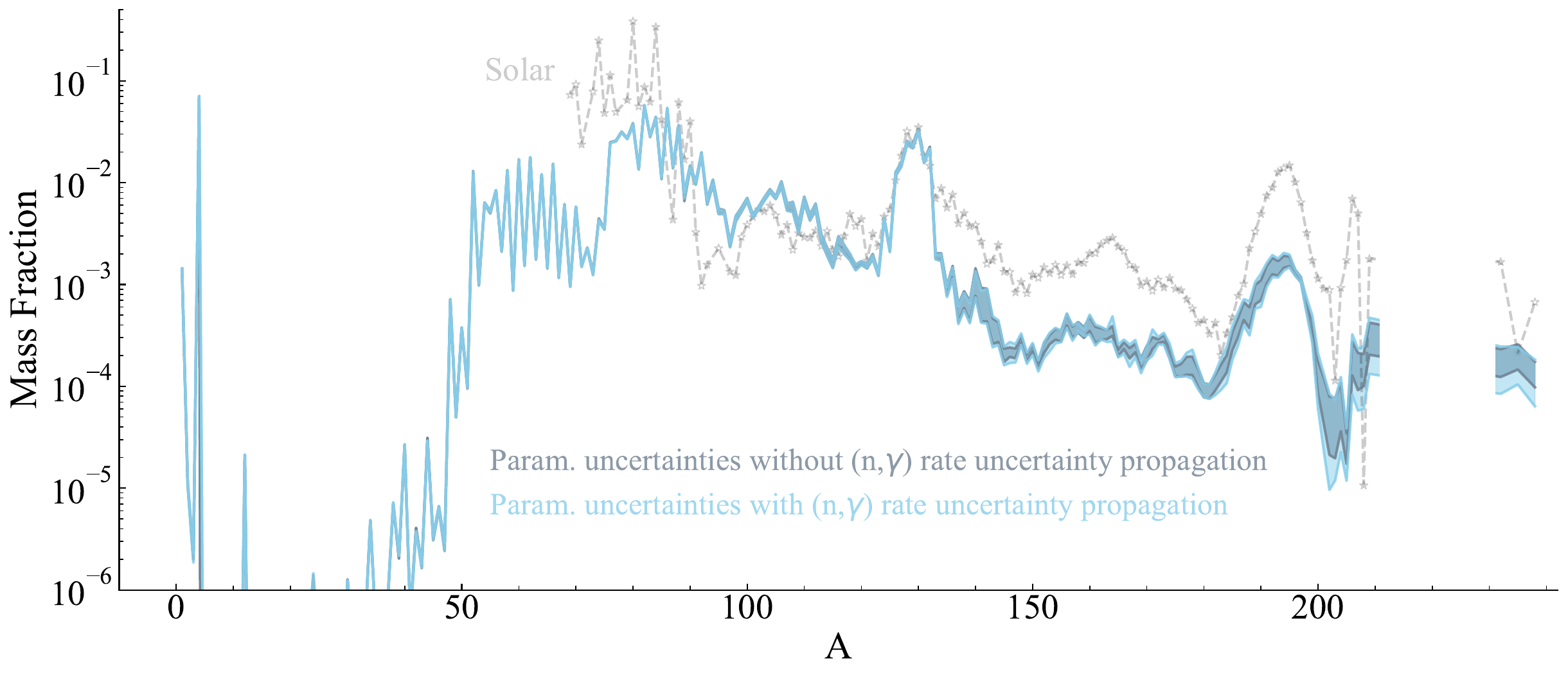}
    \caption{Same as Fig.~\ref{fig:case1_vs_4} when considering the correlated $S_n$ uncertainties with (blue results) or without (grey results) their coherent propagation into the neutron capture and photoneutron emission rates.}
    %Impact of the nuclear mass uncertainties on the mass fractions of stable nuclei (and long-lived Th and U) of the material ejected from the binary 1.375-1.375 M$_\odot$ NSM model as a function of the atomic mass A. Solar r-abundances (Goriely 1999) are shown in grey as a reference and are scaled to the peak value of the A = 130 abundance in the simulation.}
    \label{fig:ng_abundances}
\end{figure}

To this point, mass uncertainties have not been directly propagated to neutron capture rates due to the significant computational demands. However, for the case of correlated $S_n$, the limited number of parameter combinations allows for feasible calculations of reaction rates. Using the TALYS reaction code \citep{Koning23}, neutron capture rates were coherently computed based on the associated masses for this case. These rates, along with photorates derived from the masses, were then incorporated into r-process nucleosynthesis simulations.

Figure \ref{fig:ng_abundances} compares the final abundance uncertainties resulting from correlated $S_n$ mass uncertainties, with and without propagation to neutron capture rates. The results show that abundance uncertainties remain nearly identical in both scenarios, as the well-established $(n,\gamma)$–$(\gamma,n)$ equilibrium in the specific NSM model limits the impact of this propagation. However, a slight increase in uncertainties for nuclei with $A \ga 200$ is observed when neutron capture rates are included. This analysis demonstrates that the earlier approach, which propagated nuclear mass uncertainties without explicitly calculating neutron capture rates, provides a robust first-order approximation.

\section{Conclusion}

This study emphasizes the critical role of statistical parameter uncertainties in nuclear astrophysics and their impact on nucleosynthesis processes, specifically the i-process in low-metallicity AGB stars and the r-process in NSMs. Using the BFMC method, we have developed a coherent framework to quantify these uncertainties by anchoring theoretical models to experimental constraints.

We emphasize the importance of maintaining consistency within the same nuclear model and network to avoid overestimating uncertainties due to neglected correlations. The recommended approach involves using the BFMC-derived maximum and minimum limits as constraints for Monte Carlo simulations, ensuring uniform distribution of parameter variability.

For the i-process, we demonstrated how parameter uncertainties in neutron capture rates influence the predicted surface abundances of AGB stars. Odd-$Z$ elements were found to exhibit larger uncertainties due to their reliance on single isotopes, making them highly sensitive to specific reaction rates. Our analysis highlights the importance of experimentally constraining key neutron capture rates to improve nucleosynthesis predictions.

For the r-process in neutron star mergers, we examined the propagation of nuclear mass uncertainties, particularly for neutron-rich nuclei far from stability. These uncertainties, driven by unknown masses, propagate through correlated quantities such as neutron separation ($S_n$) and $\beta$-decay ($Q_\beta$) energies, affecting the prediction of ejecta compositions. The study revealed that correlated mass uncertainties result in less scatter in the abundance predictions, especially for elements beyond the second r-process peak. This emphasizes the importance of considering mass correlations in r-process simulations and the potential impact of experimental mass measurements in refining these models.

Overall, this study demonstrates that parameter uncertainties have an impact comparable to systematic model uncertainties, underscoring the need for expanded experimental data to constrain nuclear properties. Future efforts should prioritize precise measurements of key reaction rates and masses to enhance the predictive power of nucleosynthesis models and deepen our understanding of stellar evolution and the origin of heavy elements in the universe.

\label{Sect:conclusion}

% \begin{acknowledgements}
\bmhead{Acknowledgements}
SM and SG has received support from the European Union (ChECTEC-INFRA, project no. 101008324). This work was supported by the F.R.S.-FNRS under Grant No IISN 4.4502.19 and by the F.R.S.-FNRS and the FWO - Vlaanderen under the EOS Project Nr O000422. The present research benefited from computational resources made available on Lucia, the Tier-1 supercomputer of the Walloon Region, infrastructure funded by the Walloon Region under the grant agreement n°1910247. LS and SG are senior F.R.S.-FNRS research associates.
A.C. is a Postdoctoral Researcher of the F.R.S.–FNRS.
\def\aj{AJ}%
          % Astronomical Journal
\def\actaa{Acta Astron.}%
          % Acta Astronomica
\def\araa{ARA\&A}%
          % Annual Review of Astron and Astrophys
\def\apj{ApJ}%
          % Astrophysical Journal
\def\apjl{ApJ}%
          % Astrophysical Journal, Letters
\def\apjs{ApJS}%
          % Astrophysical Journal, Supplement
\def\ao{Appl.~Opt.}%
          % Applied Optics
\def\apss{Ap\&SS}%
          % Astrophysics and Space Science
\def\aap{A\&A}%
          % Astronomy and Astrophysics
\def\aapr{A\&A~Rev.}%
          % Astronomy and Astrophysics Reviews
\def\aaps{A\&AS}%
          % Astronomy and Astrophysics, Supplement
\def\azh{AZh}%
          % Astronomicheskii Zhurnal
\def\baas{BAAS}%
          % Bulletin of the AAS
\def\bac{Bull. astr. Inst. Czechosl.}%
          % Bulletin of the Astronomical Institutes of Czechoslovakia 
\def\caa{Chinese Astron. Astrophys.}%
          % Chinese Astronomy and Astrophysics
\def\cjaa{Chinese J. Astron. Astrophys.}%
          % Chinese Journal of Astronomy and Astrophysics
\def\icarus{Icarus}%
          % Icarus
\def\jcap{J. Cosmology Astropart. Phys.}%
          % Journal of Cosmology and Astroparticle Physics
\def\jrasc{JRASC}%
          % Journal of the RAS of Canada
\def\mnras{MNRAS}%
          % Monthly Notices of the RAS
\def\memras{MmRAS}%
          % Memoirs of the RAS
\def\na{New A}%
          % New Astronomy
\def\nar{New A Rev.}%
          % New Astronomy Review
\def\pasa{PASA}%
          % Publications of the Astron. Soc. of Australia
\def\pra{Phys.~Rev.~A}%
          % Physical Review A: General Physics
\def\prb{Phys.~Rev.~B}%
          % Physical Review B: Solid State
\def\prc{Phys.~Rev.~C}%
          % Physical Review C
\def\prd{Phys.~Rev.~D}%
          % Physical Review D
\def\pre{Phys.~Rev.~E}%
          % Physical Review E
\def\prl{Phys.~Rev.~Lett.}%
          % Physical Review Letters
\def\pasp{PASP}%
          % Publications of the ASP
\def\pasj{PASJ}%
          % Publications of the ASJ
\def\qjras{QJRAS}%
          % Quarterly Journal of the RAS
\def\rmxaa{Rev. Mexicana Astron. Astrofis.}%
          % Revista Mexicana de Astronomia y Astrofisica
\def\skytel{S\&T}%
          % Sky and Telescope
\def\solphys{Sol.~Phys.}%
          % Solar Physics
\def\sovast{Soviet~Ast.}%
          % Soviet Astronomy
\def\ssr{Space~Sci.~Rev.}%
          % Space Science Reviews
\def\zap{ZAp}%
          % Zeitschrift fuer Astrophysik
\def\nat{Nature}%
          % Nature
\def\iaucirc{IAU~Circ.}%
          % IAU Cirulars
\def\aplett{Astrophys.~Lett.}%
          % Astrophysics Letters
\def\apspr{Astrophys.~Space~Phys.~Res.}%
          % Astrophysics Space Physics Research
\def\bain{Bull.~Astron.~Inst.~Netherlands}%
          % Bulletin Astronomical Institute of the Netherlands
\def\fcp{Fund.~Cosmic~Phys.}%
          % Fundamental Cosmic Physics
\def\gca{Geochim.~Cosmochim.~Acta}%
          % Geochimica Cosmochimica Acta
\def\grl{Geophys.~Res.~Lett.}%
          % Geophysics Research Letters
\def\jcp{J.~Chem.~Phys.}%
          % Journal of Chemical Physics
\def\jgr{J.~Geophys.~Res.}%
          % Journal of Geophysics Research
\def\jqsrt{J.~Quant.~Spec.~Radiat.~Transf.}%
          % Journal of Quantitiative Spectroscopy and Radiative Trasfer
\def\memsai{Mem.~Soc.~Astron.~Italiana}%
          % Mem. Societa Astronomica Italiana
\def\nphysa{Nucl.~Phys.~A}%
          % Nuclear Physics A
\def\physrep{Phys.~Rep.}%
          % Physics Reports
\def\physscr{Phys.~Scr}%
          % Physica Scripta
\def\planss{Planet.~Space~Sci.}%
          % Planetary Space Science
\def\procspie{Proc.~SPIE}%
          % Proceedings of the SPIE
\let\astap=\aap
\let\apjlett=\apjl
\let\apjsupp=\apjs
\let\applopt=\ao
\bibliography{ref_npa,add_bib}% common bib file


\end{document}